\documentclass[10pt,letterpaper]{article} 
\usepackage{opex3}
\usepackage{color}
\usepackage{cite}

\def\VO2{\mathrm{VO_2}}
\def\U#1{\,\mathrm{#1}}

\begin{document}
%
%
\title{Dynamically Babinet-invertible metasurface: a capacitive-inductive reconfigurable filter for terahertz waves using vanadium-dioxide metal-insulator transition}

\author{Yoshiro Urade,$^{1,*}$ Yosuke Nakata,$^2$ Kunio Okimura,$^3$ Toshihiro Nakanishi,$^1$ Fumiaki Miyamaru,$^{2,4}$ Mitsuo W. Takeda,$^4$ and Masao Kitano$^1$}

\address{$^1$Department of Electronic Science and Engineering, Kyoto University, Kyoto 615-8510, Japan\\
$^2$Center for Energy and Environmental Science, Shinshu University, 4-17-1 Wakasato, Nagano 380-8553, Japan\\
$^3$School of Engineering, Tokai University, 4-1-1 Kitakaname, Hiratsuka, Kanagawa 259-1292, Japan\\
$^4$Department of Physics, Shinshu University, Nagano 390-8621, Japan}

\email{$^*$urade@giga.kuee.kyoto-u.ac.jp} 



\begin{abstract}
 This paper proposes a reconfigurable planar metamaterial that can be switched between capacitive and inductive responses using local changes in the electrical conductivity of its constituent material.
 The proposed device is based on Babinet's principle and exploits the singular electromagnetic responses of metallic checkerboard structures, which are dependent on the local electrical conductivity.
 Utilizing the heating-induced metal-insulator transition of vanadium dioxide~($\VO2$), the proposed metamaterial is designed to compensate for the effect of the substrate and is experimentally characterized in the terahertz regime.
This reconfigurable metamaterial can be utilized as a switchable filter and as a switchable phase shifter for terahertz waves.
\end{abstract}


\bibliographystyle{osajnl}

\section{Introduction}
Metamaterials are artificial electromagnetic media that are comprised of unit structures smaller than the wavelength of interest.
Their responses can be tailored by tuning the constituent materials, sizes, shapes, and arrangement of the unit cells.
Two-dimensional metamaterials are called metasurfaces.
Recently, considerable efforts have been made towards the realization of tunable or reconfigurable metamaterials~\cite{Zheludev2012metadevice}, the properties of which are controllable using external stimuli, such as electric fields~\cite{Kan2013}, light~\cite{Shadrivov2012}, and temperature ($T$) changes~\cite{Ou2011}.

Metamaterial and tunable metamaterial technology is especially important with regard to the terahertz region of the electromagnetic spectrum, because this frequency region lacks efficient manipulation techniques based on conventional electronics or photonics~\cite{Tao2011}.
To eliminate this shortcoming, various active and passive devices based on the metamaterial concept have been studied and developed, including amplitude modulators~\cite{Chen2006}, phase modulators~\cite{Chen2009}, group delay modulators~\cite{Miyamaru2014}, and thin absorbers~\cite{Tao2008}.

In this paper, we propose a reconfigurable metasurface that can alternate between two contrasting behaviors in response to local changes in the electrical conductivity of its constituent material.
The developed metasurface is based on metallic-checkerboard structural singularities~\cite{Edmunds2010,Kempa2006,Takano2014,gonzalez2015basic,Nakata2013,Urade2015}, the electromagnetic responses of which are strongly dependent on the interconnection state of the metallic squares of the checkerboard.
For example, these structures exhibit a high-pass frequency characteristic with a transmission peak for propagating incident waves when the metallic squares are electrically interconnected~(``inductive'' behavior). However,  a low-pass frequency characteristic with a dip is obtained if the metallic parts are not interconnected~(``capacitive'' behavior)~\cite{Takano2014}.
Therefore, we can dynamically alter the behavior of these devices by switching the local electrical conductivity at the interconnection points of the metallic squares.
This reconfigurable characteristic of the metasurface can be exploited for switchable filter and switchable phase shifter applications for terahertz waves.
We note that, although similar functionality has been realized by exploiting the characteristics of anisotropic self-complementary metasurfaces~\cite{Ortiz2013}, mechanical rotation of the metasurface itself is required to accomplish the reconfiguration in such cases.

Here, we use vanadium dioxide~($\VO2$) to achieve the electrical conductivity switching.
$\VO2$ exhibits a metal-insulator transition at $T$ $\approx$ $340\U{K}$~\cite{Morin1959}, which is relatively close to room temperature, and the electrical conductivity typically changes by several orders of magnitude.
In addition, the $\VO2$ phase transition can be triggered by other external stimuli, including a high electric field~\cite{Okimura2006} and photoexcitation~\cite{Cavalleri2001}.
Hence, $\VO2$ is a suitable material for realization of the proposed reconfigurable metasurface.


\section{Operating principles}

Babinet's principle for vector waves relates the scattering problem for a metallic metasurface to that of its complementary structure, which is obtained by interchanging the metallic and vacant areas of the original metasurface~\cite{kong1990electromagnetic}.
This principle can be extended to cases where the metallic metasurfaces contain resistive elements~\cite{Nakata2013,BaumNote1974}.
In this extension, regions with high electrical conductivity have corresponding low-electrical-conductivity regions in the complementary structure, and vice versa~(this operation is called \textit{Babinet inversion}).
Hereafter, we assume that the incident wave is a linearly polarized plane wave with angular frequency $\omega$.
Note that the incident polarization in the complementary scattering problem is orthogonal to that in the original problem.
For the zeroth-order transmission mode, a simple relationship exists between the original and complementary problems~\cite{Urade2015, Nakata2013}

\begin{equation}
 \tilde{t}(\omega)+\tilde{t}_\mathrm{c}(\omega)=1,
  \label{eq:Babinet}
\end{equation}
where $\tilde{t}(\omega)$ and $\tilde{t}_\mathrm{c}(\omega)$ are the complex amplitude transmission coefficients of the zeroth-order transmission modes in the original problem and its complement, respectively.
Moreover, if scattering into higher-order diffraction modes and ohmic losses are negligible in each problem, $\tilde{t}(\omega)$ and $\tilde{t}_\mathrm{c}(\omega)$ are in quadrature phase to each other~\cite{Baena2015}.

\begin{figure}[htb]
\centering\includegraphics[width=10.5cm]{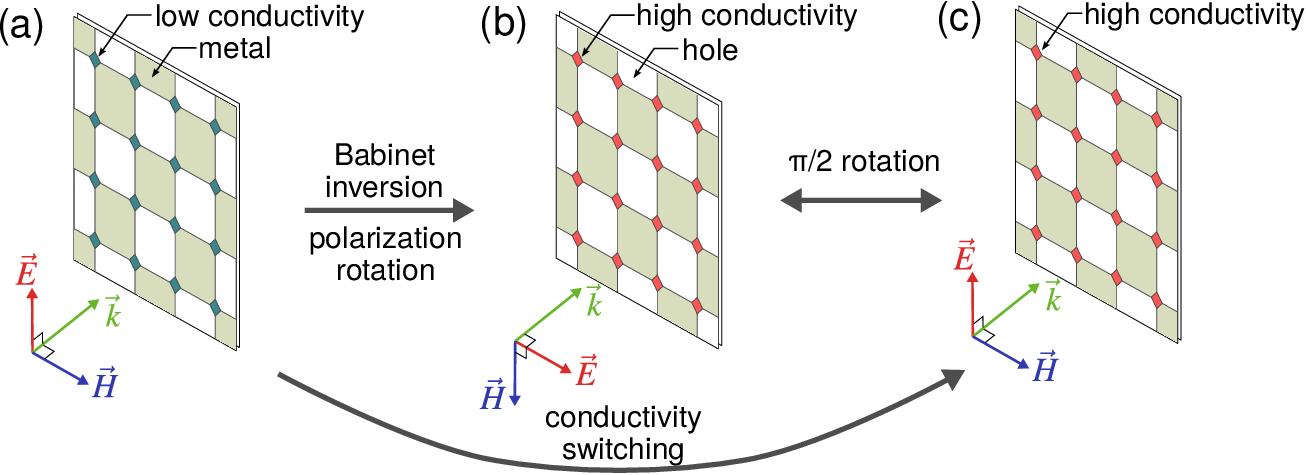}
\caption{Principles of Babinet-invertible metasurface. (a) Off state. (b) Babinet-inverted structure. Note that the polarization of the incident wave is orthogonal to that in the off state. (c) On state. The vectors $\vec{E}$, $\vec{H}$, and $\vec{k}$ indicate the electric field, the magnetic field, and the wavevector of the incident plane waves, respectively.}
 \label{fig:concept}
\end{figure}

Let us consider a plane-wave scattering problem involving a metallic checkerboard-like metasurface, as shown in Fig.~\ref{fig:concept}(a).
We assume that the local electrical conductivity at the interconnection points of the metallic patches is sufficiently low and, thus, the metallic squares are electrically disconnected.
Such a structure exhibits capacitive behavior with a transmission dip at frequency $\omega_0$~[$\tilde{t}(\omega_0)\simeq 0$].
This is referred to as the \textit{off} state.
The corresponding Babinet-inverted structure is shown in Fig.~\ref{fig:concept}(b).
In this case, as a result of Babinet's principle, the local electrical conductivity at the interconnection points of the metallic patches must be high, and the polarization of the incident plane wave is rotated by $\pi/2$.
The Babinet-inverted structure exhibits inductive behavior with a transmission peak at $\omega_0$, according to Babinet's principle~[$\tilde{t}_\mathrm{c}(\omega_0)\simeq 1$].
However, if the electrical conductivity at the interconnection points is increased, the off state in Fig.~\ref{fig:concept}(a) transits to the \textit{on} state shown in Fig.~\ref{fig:concept}(c).
In other words, the metallic squares become electrically connected.
As a result of the four-fold rotational symmetry of the metasurface, the scattering problem of the on state shown in Fig.~\ref{fig:concept}(c) is equivalent to that of the Babinet-inverted structure shown in Fig.~\ref{fig:concept}(b), provided normal incidence is assumed.
Therefore, the transmission characteristic in the on state is identical to that in the Babinet-inverted structure derived from the off state.
This means that we can dynamically alter the behavior of the metasurface to the complementary behavior of its Babinet-inverted structure, simply by switching the local electrical conductivity at the interconnection points of the checkerboard.
Hence, we refer to this reconfigurable metasurface as a \textit{Babinet-invertible metasurface}.

It should be noted that Babinet's principle does not hold strictly if the planar structure is placed on a substrate, which breaks the inversion symmetry with respect to the metasurface plane.
However, it is known that a qualitatively similar relationship exists between the characteristics of the metallic complementary structures, even in the presence of substrates~\cite{Chen:07,Zentgraf2007}.
Therefore, the above principle is applicable to a Babinet-invertible metasurface with moderate-refractive-index substrates.

\section{Sample design and fabrication}
\begin{figure}[htb]
\centering\includegraphics[height=3.6cm]{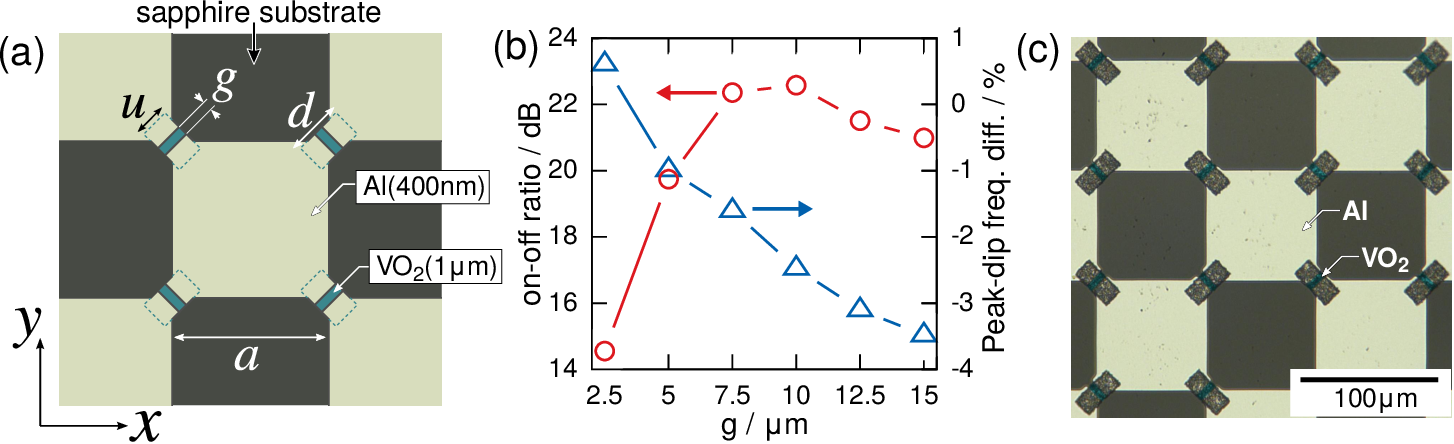}
\caption{(a)~ Babinet-invertible metasurface design. The parameters are as follows: $a=75\U{\mu m}$, $g=5\U{\mu m}$, $u=15\U{\mu m}$, and $d=32\U{\mu m}$. The broken lines indicate the area where the Al structures overlap the $\VO2$. (b)~Calculated $g$ dependence of the on-off ratio of the power transmission at the peak and dip frequencies and their frequency difference. (c)~Photograph of fabricated metasurface. The exposed regions of the $\VO2$ patches are digitally colored in turquoise for clarity.}
  \label{fig:design}
\end{figure}

Figure~\ref{fig:design}(a) shows the design employed to realize the Babinet-invertible metasurface using $\VO2$.
The small $\VO2$ patches are located at the interconnection points of the checkerboard.
Below the critical temperature $T_\mathrm{c}\sim 340\U{K}$ of $\VO2$, the patches are insulating and, therefore, the metasurface is in the off state. By increasing $T$ above $T_\mathrm{c}$, the $\VO2$ insulator-to-metal transition is induced and, thus, the metasurface transits to the on state.
The Al structures overlap the $\VO2$ patches, ensuring electrical connection. 
The width $g$ of the gaps between the metallic structures in Fig.~\ref{fig:design}(a) is narrower than those in the ideal case without a substrate, as shown in Fig.~\ref{fig:concept}.
This modification is introduced in order to compensate for the difference between the dip and peak frequencies in the transmission characteristics due to the dielectric substrate. This compensation is achieved by increasing the gap capacitance; otherwise, the transmission dip frequency would be slightly higher than the transmission peak frequency.
In addition, the small $g$ decreases the resistance of the interconnection points, which leads to a high transmission peak in the on state.
The optimal value of $g$ is heuristically found via numerical simulations using a commercial finite-element method solution package~(COMSOL Multiphysics). Figure~\ref{fig:design}(b) shows the calculated $g$ dependence of the on-off ratio of the power transmission at the peak and dip frequencies and their frequency difference. To obtain high on-off ratio and small frequency difference, we choose $g=5\U{\mu m}$.

The designed metasurface is fabricated as follows.
A thin film of stoichiometric $\VO2$ with 1-$\mu$m thickness is prepared on a sapphire $(0001)$ substrate~(thickness: $0.9~\U{mm}$) via reactive magnetron sputtering of a vanadium target~\cite{Okimura2005}. The sputtering conditions are as follows:
substrate $T$: $400\U{^\circ C}$; 
radio frequency (rf) power: $200\U{W}$; 
Ar flow rate: $0.11\U{Pa\cdot m^3/s}$;
$\mathrm{O_2}$ flow rate: $1.4\times 10^{-3}\U{Pa\cdot m^3/s}$; and
total Ar and $\mathrm{O_2}$ pressure: $0.5\U{Pa}$. These conditions yield
a deposition rate of $\sim8\U{nm/min}$.
Photoresist is spin-coated on the $\VO2$ film and then patterned as an etching mask using photolithography.
The exposed part of the $\VO2$ film is removed by wet etching with a mixture comprised of phosphoric acid, nitric acid, acetic acid, and water, at a 16:1:2:1 ratio.
Then, an Al checkerboard~(thickness:~$400\U{nm}$) is fabricated on the sample via standard photolithography, electron beam evaporation at room temperature, and lift-off techniques.
A photograph of the sample is shown in Fig.~\ref{fig:design}(c).
As a result of the surface roughness of the $\VO2$ patches, the regions in which the Al structures overlap the $\VO2$ appear dark and roughly textured compared with the other Al regions.

\section{Characterization}
We use the terahertz time-domain spectroscopy technique~\cite{Grischkowsky1990} for characterization of the fabricated metasurface.
The metasurface is set on a sample holder, the temperature of which is stabilized through temperature feedback control of the electric current in a nichrome wire attached to the holder.
The value of the holder $T$ is monitored using a thermocouple, and
the sample holder is covered by styrene boards for thermal insulation.
We generate and detect terahertz pulses through the photoexcitation of photoconductive dipole antennas on low-temperature-grown GaAs substrates.
The emitted terahertz beam is linearly polarized and focused onto the sample under normal incidence by a Si hyper-hemispherical lens and a subsequent Tsurupica lens. The beam diameter at the focal plane is $\sim 5\U{mm}$ and the
incident polarization is in the $y$-direction indicated in Fig.~\ref{fig:design}(a).
While changing the $T$ value of the sample holder, we measure the temporal waveforms of the terahertz electric fields transmitted through the sample and, also, through a plain sapphire plate (as a reference measurement).
These measurements are performed under dry air purged conditions.
The transmission coefficient spectra $\tilde{t}(\omega)$ are obtained by dividing the Fourier transforms of the electric fields transmitted through the sample by those of the reference measurement, where the echo pulses caused by the reflections at the sample-air boundaries are removed by setting a time window, and the time dependence $\exp (\mathrm{j} \omega t)$ is assumed.

\begin{figure}[htb]
\centering\includegraphics[width=12.5cm]{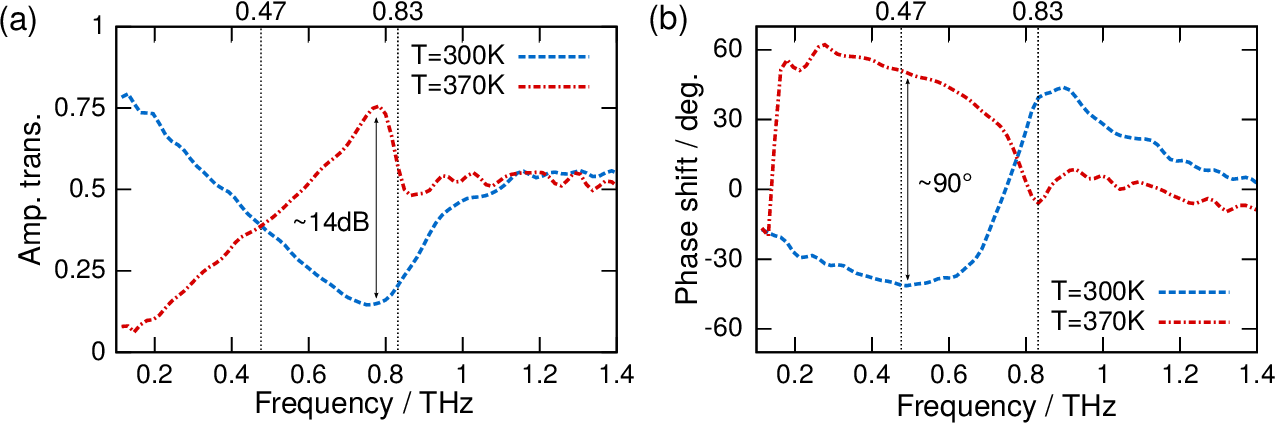}
\caption{Measured transmission coefficient spectra at $T=300$ and $370\U{K}$: (a)~amplitude transmission spectra and (b)~phase shift spectra.}
  \label{fig:result}
\end{figure}

Figures~\ref{fig:result}(a) and \ref{fig:result}(b) show the measured transmission amplitude $|\tilde{t}(\omega)|$ and phase shift $\arg\tilde{t}(\omega)$ spectra, respectively, at $T=300$ and $370\U{K}$.
Hence, it is confirmed that the response of the metasurface switches from capacitive to inductive behavior with increased $T$.
In other words, the off state at low $T$ transits to the on state above the $T_c$ of the $\VO2$.
It should be noted that the metasurface in the on state returns to the off state if it cools down due to the reversibility of the $\VO2$ phase transition.
At $0.77\U{THz}$, the modulation depth amounts to $14\U{dB}$, which is limited by the conductivities of the insulating and metallic $\VO2$ phases at low and high $T$, respectively.
In the lower-frequency region of the phase spectra much below the lowest diffraction frequency~$0.83\U{THz}$, where scattering into higher-order diffraction modes is negligible, there is an almost constant phase difference of $\sim\pi/2$ between $T=300$ and $370\U{K}$, as a result of Babinet's principle.
In particular, at $0.47\U{THz}$ or the crossing frequency of the two amplitude spectra, the relative transmission phase shift of the metasurface can be switched to the quadrature phase without changing the transmission amplitude.
We note that these characteristics of the metasurface are independent of the incident polarization states, provided normal incidence is assumed; this behavior is a result of the four-fold rotational symmetry.

\section{Conclusion}
In conclusion, we have proposed a reconfigurable metasurface that is transformable into its Babinet-inverted structure by exploiting the singularities of a metallic checkerboard structure and the metal-insulator transition of $\VO2$. The operation of the metasurface has been confirmed experimentally.
The Babinet-invertible metasurface equipped with a thermostat can be utilized as a switchable filter and as a switchable $\pi/2$ phase shifter for terahertz waves.
The design principles discussed in this study can be realized using not only $\VO2$, but also other conductivity-tunable materials or structures such as photodoped semiconductors, superconductors, non-volatile conductivity changes used in resistive memory devices, and microelectromechanical systems.
We also note that the present metasurface is expected to exhibit nonlinear behavior for intense terahertz pulses inducing phase transition of the $\VO2$, which will be used as an optical fuse~\cite{Strikwerda2015}. 

\section*{Acknowledgments}
The authors thank Dr.~Keisuke Takano and Prof.~Masanori Hangyo for useful discussions, and Dr.~Joe Sakai for his helpful comments on the etching process.
This work was partly supported by JSPS KAKENHI grants~(15J07603, 25790065, 25287101) and by research grants from the Murata Science Foundation~(Y.N.). The sample fabrication was performed with the help of Kyoto University Nano Technology Hub, as part of the ``Nanotechnology Platform Project'' sponsored by MEXT in Japan. One of the authors~(Y.U.) was supported by a JSPS Research Fellowship for Young Scientists.
\end{document}